\documentstyle[12pt,qqa4lart]{article}
\input tcilatex

\QQQ{Language}{
American English
}

\begin{document}

\title{Decoherence of quantum registers}
\author{Lu-Ming Duan and Guang-Can Guo\thanks{%
gcguo@sunlx06.nsc.ustc.edu.cn} \\
Department of Physics, University of Sceience \\
and Technology of China, Hefei, 230026, P.R.China}
\date{}
\maketitle

\begin{abstract}
\baselineskip 20ptWe consider decoherence of quantum registers, which
consist of the qubits sited approximately periodically in space. The sites
of the qubits are permitted to have a small random variance. We derive the
explicit conditions under which the qubits can be assumed decohering
independently. In other circumstances, the qubits are decohered
cooperatively. We describe two kinds of collective decoherence. In each
case, a scheme is proposed for reducing the collective decoherence. The
schemes operate by encoding the input states of the qubits into some
''subdecoherent'' states.\\


\newpage\ 
\end{abstract}

\baselineskip 20pt

\section{Introduction}

Quantum computation has become an active field since Shor discovered that
quantum computers could solve the problem of finding factors of a large
number in a time which is a polynomial function of the length (number of the
bits) of the number [1,2]. However, there are some obstacles to realize
quantum computation. The main one is decoherence of the qubits caused by the
interaction with the environment [3-6]. Unruh analyzed decoherence in
quantum memory with the assumption that the qubits are decohered
independently [3]. To reduce this kind of decoherence, Shor proposed a
subtle strategy called quantum error correction which could restore useful
information from the decohered states [7]. Many quantum error-correcting
codes have since been discovered to correct quantum errors occurring during
the store of the information or during the gate operations [8-24]. Apart
from the independent decoherence, there are other circumstances. The qubits
may be decohered collectively. The collective decoherence has some new
features, which make the strategy for reducing this kind of decoherence is
much different from the quantum error correction schemes [25-27].

In this paper, we consider a practical model of the quantum register. The
register consists of the qubits sited approximately periodically in space.
But the sites of the qubits are permitted to have a small random variance.
This small disorder may be due to the limited manufacture precision or
caused by the thermal variation of the qubits. Starting from a general
decoherence model of the quantum register, we obtain its exact solution.
Then we discuss in which circumstances the qubits can be regarded decohering
independently. From the cooperative decoherence to the independent
decoherence, the small disorder of the sites of the qubits plays an
important role. The independent decoherence is an ideal case. There is
another ideal case, i.e., the collective decoherence. We derive two kinds of
collective decoherence. The first case has been discussed in [25] and [26].
In this case the qubits lie in the coherent length of the environment. The
second case of the collective decoherence is new. It results from the
approximate periodicity of the register. The existing quantum error
correction schemes are not suitable for reducing the collective decoherence.
So in each case of the collective decoherence, we propose an alternate
decoherence-reducing strategy, which exploits the new feature of the
collective decoherence.

The paper is arranged as follows: The general decoherence model of the
quantum register is described and solved in Sec. 1. In Sec. 2, we derive the
explicit conditions under which the qubits can be assumed decohering
independently. There are two circumstances. Section 3 describes two kinds of
collective decoherence and the corresponding decoherence-reducing strategy.

\section{The decoherence model of quantum registers and its exact solution}

For a practical quantum register, it is reasonable to assume that the qubits
are arranged approximately periodically in space. So the coordinate of the $%
\overrightarrow{l}$ qubit can be expressed as $\overrightarrow{r}_{%
\overrightarrow{l}}=\overrightarrow{R}_{\overrightarrow{l}}+\overrightarrow{%
\delta }_{\overrightarrow{l}}$, where $\overrightarrow{R}_{\overrightarrow{l}%
}$ is a rigorous periodical function of $\overrightarrow{l}$ with a lattice
constant $d$. ( For simplicity we assume the lattice constants are same
along different directions.) $\overrightarrow{\delta }_{\overrightarrow{l}}$
is a small random variable, which satisfies $\left\langle \overrightarrow{%
\delta }_{\overrightarrow{l}}\right\rangle =0$, and $\sqrt{\left\langle 
\overrightarrow{\delta }_{\overrightarrow{l}}\cdot \overrightarrow{\delta }_{%
\overrightarrow{l}}\right\rangle }=\delta $. Generally $\delta <<d$. .
Decoherence of the qubits is caused by the coupling with the environment.
The noise field in the environment may be a radiative field ( such as
two-level atoms in a cavity [28]) or a phonon field ( such as the trapped
ions [29] ). Here we consider the decoherence by its narrow meaning, i.e.,
we only consider the dephasing process. The loss of the energy is not
included. The qubits can always be described by the Pauli operators $%
\overrightarrow{\sigma }_{\overrightarrow{l}}$ and the environment is
modelled by a bath of oscillators. The total Hamiltonian describing the
dephasing process takes the form 
\begin{equation}
\label{1}H=\hbar \left[ \omega _0\stackunder{\overrightarrow{l}}{\sum }%
\sigma _{\overrightarrow{l}}^z+\stackunder{\overrightarrow{k}}{\sum }\omega
_{\overrightarrow{k}}a_{\overrightarrow{k}}^{+}a_{\overrightarrow{k}}+%
\stackunder{\overrightarrow{k},\overrightarrow{l}}{\sum }\left( g_{%
\overrightarrow{k}\overrightarrow{l}}a_{\overrightarrow{k}}+g_{%
\overrightarrow{k}\overrightarrow{l}}^{*}a_{\overrightarrow{k}}^{+}\right)
\sigma _{\overrightarrow{l}}^z\right] ,
\end{equation}
where $a_{\overrightarrow{k}}$ is the annihilation operator of the bath mode 
$\overrightarrow{k}$ and $g_{\overrightarrow{k}\overrightarrow{l}}$ is the
coupling coefficient. If the mode functions of the noise field are plane
waves, $g_{\overrightarrow{k}\overrightarrow{l}}$ can be expressed as 
\begin{equation}
\label{2}g_{\overrightarrow{k}\overrightarrow{l}}=g_{\overrightarrow{k}}e^{-i%
\overrightarrow{k}\cdot \overrightarrow{r}_{\overrightarrow{l}}}.
\end{equation}
In the following we assume Eq. (2) holds.

We solve the decoherence model in the interaction picture. The interaction
Hamiltonian is 
\begin{equation}
\label{3}H=\hbar \stackunder{\overrightarrow{k},\overrightarrow{l}}{\sum }%
\left[ g_{\overrightarrow{k}\overrightarrow{l}}a_{\overrightarrow{k}%
}e^{-i\omega _{\overrightarrow{k}}t}+g_{\overrightarrow{k}\overrightarrow{l}%
}^{*}a_{\overrightarrow{k}}^{+}e^{i\omega _{\overrightarrow{k}}t}\right] .
\end{equation}
In Ref. [25], the time evolution operator is expressed as $U\left( t\right)
=\exp \left[ -\frac i\hbar \int_0^tH_I\left( t^{^{\prime }}\right)
dt^{^{\prime }}\right] $. But this expression is not correct since $\left[
H_I\left( t\right) ,H_I\left( t^{^{\prime }}\right) \right] \neq 0$. In fact
it is not difficult to verify that the evolution operator corresponding the
Hamiltonian (3) has the form 
\begin{equation}
\label{4}U\left( t\right) =\exp \left\{ \stackunder{\overrightarrow{k},%
\overrightarrow{l}}{\sum }\left( \xi _{\overrightarrow{k}\overrightarrow{l}%
}^{*}\left( t\right) a_{\overrightarrow{k}}^{+}-\xi _{\overrightarrow{k}%
\overrightarrow{l}}\left( t\right) a_{\overrightarrow{k}}\right) \sigma _{%
\overrightarrow{l}}^z\right\} e^{if\left( t\right) },
\end{equation}
where 
\begin{equation}
\label{5}\xi _{\overrightarrow{k}\overrightarrow{l}}\left( t\right) =\frac{%
g_{\overrightarrow{k}\overrightarrow{l}}\left( 1-e^{-i\omega _{%
\overrightarrow{k}}t}\right) }{\omega _{\overrightarrow{k}}},
\end{equation}
and 
\begin{equation}
\label{6}f\left( t\right) =\stackunder{\overrightarrow{k}}{\sum }\left\{ 
\frac{\omega _{\overrightarrow{k}}t-\sin \left( \omega _{\overrightarrow{k}%
}t\right) }{\omega _{\overrightarrow{k}}^2}\cdot \left| \stackunder{%
\overrightarrow{l}}{\sum }\left( g_{\overrightarrow{k}\overrightarrow{l}%
}\sigma _{\overrightarrow{l}}^z\right) \right| ^2\right\} .
\end{equation}
In the following, we will see the factor $e^{if\left( t\right) }$ in Eq. (4)
missed by Ref. [25] results in the Lamb phase shift, which plays an
important role in the collective decoherence.

The time evolution of the register is completely determined by the operator $%
U\left( t\right) $. To see this, let $\rho _{i_{\overrightarrow{l}},j_{%
\overrightarrow{l}}}=\left| i_{\overrightarrow{l}}\right\rangle \left\langle
j_{\overrightarrow{l}}\right| $ and 
\begin{equation}
\label{7}\rho _{\left\{ i_{\overrightarrow{l}},j_{\overrightarrow{l}%
}\right\} }=\rho _{i_1,j_1}\otimes \rho _{i_2,j_2}\otimes \cdots \otimes
\rho _{i_L,j_L},
\end{equation}
where $i_{\overrightarrow{l}}=\pm 1,$ $j_{\overrightarrow{l}}=\pm 1$, and $%
\left| \pm 1\right\rangle $ are two eigenstates of the operator $\sigma ^z$. 
$L=L_1L_2L_3$ is the total number of the qubits. With this notation, the
initial density $\rho _s\left( 0\right) $ of the register can be expanded
into 
\begin{equation}
\label{8}\rho _s\left( 0\right) =\stackunder{\left\{ i_{\overrightarrow{l}%
},j_{\overrightarrow{l}}\right\} }{\sum }c_{\left\{ i_{\overrightarrow{l}%
},j_{\overrightarrow{l}}\right\} }\rho _{\left\{ i_{\overrightarrow{l}},j_{%
\overrightarrow{l}}\right\} }.
\end{equation}
The environment is supposed in the thermal equilibrium. So its initial
density in the coherent representation has the form [30] 
\begin{equation}
\label{9}\rho _{env}\left( 0\right) =\stackunder{\overrightarrow{k}}{\prod }%
\int d^2\alpha _{\overrightarrow{k}}\frac 1{\pi \left\langle N_{\omega _{%
\overrightarrow{k}}}\right\rangle }\exp \left( -\frac{\left| \alpha _{%
\overrightarrow{k}}\right| ^2}{\left\langle N_{\omega _{\overrightarrow{k}%
}}\right\rangle }\right) \left| \alpha _{\overrightarrow{k}}\right\rangle
\left\langle \alpha _{\overrightarrow{k}}\right| ,
\end{equation}
where $\left\langle N_{\omega _{\overrightarrow{k}}}\right\rangle $ is the
mean photon or phonon number of the mode $\overrightarrow{k}$%
\begin{equation}
\label{10}\left\langle N_{\omega _{\overrightarrow{k}}}\right\rangle =\frac
1{\exp \left( \frac{\hbar \omega _{\overrightarrow{k}}}{k_BT}\right) -1}.
\end{equation}
When the operator (4) acts on the coherent state $\left| \alpha _{%
\overrightarrow{k}}\right\rangle $, it only generates a displacement. So
with this evolution operator, the reduced density of the register at time $t$
can easily be obtained. We have 
\begin{equation}
\label{11}
\begin{array}{c}
\rho _s\left( t\right) =
\stackunder{\left\{ i_{\overrightarrow{l}},j_{\overrightarrow{l}}\right\} }{%
\sum }c_{\left\{ i_{\overrightarrow{l}},j_{\overrightarrow{l}}\right\} }tr_{%
\overrightarrow{k}}\left[ U\left( t\right) \rho _{\left\{ i_{\overrightarrow{%
l}},j_{\overrightarrow{l}}\right\} }\otimes \rho _{env}\left( 0\right)
U^{-1}\left( t\right) \right]  \\  \\ 
=\stackunder{\left\{ i_{\overrightarrow{l}},j_{\overrightarrow{l}}\right\} }{%
\sum }c_{\left\{ i_{\overrightarrow{l}},j_{\overrightarrow{l}}\right\} }\rho
_{\left\{ i_{\overrightarrow{l}},j_{\overrightarrow{l}}\right\} }\exp \left[
-\eta _{\left\{ i_{\overrightarrow{l}},j_{\overrightarrow{l}}\right\}
}\left( t\right) +i\phi _{\left\{ i_{\overrightarrow{l}},j_{\overrightarrow{l%
}}\right\} }\left( t\right) \right] ,
\end{array}
\end{equation}
where the phase damping factor 
\begin{equation}
\label{12}\eta _{\left\{ i_{\overrightarrow{l}},j_{\overrightarrow{l}%
}\right\} }\left( t\right) =\stackunder{\overrightarrow{k}}{\sum }\left| g_{%
\overrightarrow{k}}\right| ^2\coth \left( \frac{\hbar \omega _{%
\overrightarrow{k}}}{2k_BT}\right) \frac{1-\cos \left( \omega _{%
\overrightarrow{k}}t\right) }{\omega _{\overrightarrow{k}}^2}\lambda _{1%
\overrightarrow{k}},
\end{equation}
and the Lamb phase shift 
\begin{equation}
\label{13}\phi _{\left\{ i_{\overrightarrow{l}},j_{\overrightarrow{l}%
}\right\} }\left( t\right) =\stackunder{\overrightarrow{k}}{\sum }\left| g_{%
\overrightarrow{k}}\right| ^2\cdot \frac{\omega _{\overrightarrow{k}}t-\sin
\left( \omega _{\overrightarrow{k}}t\right) }{\omega _{\overrightarrow{k}}^2}%
\lambda _{2\overrightarrow{k}}
\end{equation}
In Eqs. (12) and (13), $\lambda _{1\overrightarrow{k}}$ and $\lambda _{2%
\overrightarrow{k}}$ are defined as follows 
\begin{equation}
\label{14}\lambda _{1\overrightarrow{k}}=\left| \stackunder{\overrightarrow{l%
}}{\sum }\left( i_{\overrightarrow{l}}-j_{\overrightarrow{l}}\right) e^{i%
\overrightarrow{k}\cdot \overrightarrow{r}_{\overrightarrow{l}}}\right| ^2,
\end{equation}
\begin{equation}
\label{15}\lambda _{2\overrightarrow{k}}=\left| \stackunder{\overrightarrow{l%
}}{\sum }i_{\overrightarrow{l}}e^{i\overrightarrow{k}\cdot \overrightarrow{r}%
_{\overrightarrow{l}}}\right| ^2-\left| \stackunder{\overrightarrow{l}}{\sum 
}j_{\overrightarrow{l}}e^{i\overrightarrow{k}\cdot \overrightarrow{r}_{%
\overrightarrow{l}}}\right| ^2.
\end{equation}
In the derivation of Eqs. (12) and (13), the decomposition (2) of the
coupling coefficient has been used.

It is convenient to use the state fidelity to describe the decoherence. For
a pure input state $\left| \Psi \left( 0\right) \right\rangle $, the
fidelity is defined as 
\begin{equation}
\label{16}F=\left\langle \Psi \left( 0\right) \right| \rho _s\left( t\right)
\left| \Psi \left( 0\right) \right\rangle .
\end{equation}
Suppose the input state of the register is pure and expressed as $\left|
\Psi \left( 0\right) \right\rangle =\stackunder{\left\{ i_{\overrightarrow{l}%
}\right\} }{\sum }c_{\left\{ i_{\overrightarrow{l}}\right\} }\left| \left\{
i_{\overrightarrow{l}}\right\} \right\rangle $, then from Eq. (11) the
fidelity is 
\begin{equation}
\label{17}F=\stackunder{\left\{ i_{\overrightarrow{l}},j_{\overrightarrow{l}%
}\right\} }{\sum }\left| c_{\left\{ i_{\overrightarrow{l}}\right\} }\right|
^2\left| c_{\left\{ j_{\overrightarrow{l}}\right\} }\right| ^2\exp \left[
-\eta _{\left\{ i_{\overrightarrow{l}},j_{\overrightarrow{l}}\right\}
}\left( t\right) +i\phi _{\left\{ i_{\overrightarrow{l}},j_{\overrightarrow{l%
}}\right\} }\left( t\right) \right] .
\end{equation}
From this expression, we see that the phase damping and the Lamb phase shift
all contribute to the decoherence of the state. Eq. (13) reveals that the
phase shift increases with time approximately linearly. So with a sufficient
large $t$ the phase shift will play an important role. In the next section
we will show the factor $\lambda _{2\overrightarrow{k}}$ reduces to zero for
the independent decoherence. Therefore, the Lamb phase shift only
contributes to the cooperative decoherence.

The two factors $\lambda _{1\overrightarrow{k}}$ and $\lambda _{2%
\overrightarrow{k}}$ defined by (14) and (15) are important in determining
whether the qubits are decohered independently or collectively. We discuss
this problem in the following two sections.

\section{Independent decoherence}

We first look at the phase damping. Eq. (12) can be rewritten as 
\begin{equation}
\label{18}\eta _{\left\{ i_{\overrightarrow{l}},j_{\overrightarrow{l}%
}\right\} }\left( t\right) =x\stackunder{\overrightarrow{k}}{\sum }h_1\left(
\omega _{\overrightarrow{k}}\right) \lambda _{1\overrightarrow{k}},
\end{equation}
where $h_1\left( \omega _{\overrightarrow{k}}\right) $ is a normalized
distribution which satisfies $\stackunder{\overrightarrow{k}}{\sum }%
h_1\left( \omega _{\overrightarrow{k}}\right) =1$. $x$ is the normalization
constant%
$$
x=\stackunder{\overrightarrow{k}}{\sum }\left| g_{\overrightarrow{k}}\right|
^2\coth \left( \frac{\hbar \omega _{\overrightarrow{k}}}{2k_BT}\right) \frac{%
1-\cos \left( \omega _{\overrightarrow{k}}t\right) }{\omega _{%
\overrightarrow{k}}^2} 
$$
The expression of $h_1\left( \omega _{\overrightarrow{k}}\right) $ is given
by comparing (18) with (12). Its explicit form depends on the coupling
coefficient $\left| g_{\overrightarrow{k}}\right| ^2$, whereas the latter is
determined by the specific characteristics of the physical system. But here
we take a simplification. The distribution $h_1\left( \omega _{%
\overrightarrow{k}}\right) $ is approximately characterized by its mean
value $\overline{\omega }_1$ and variance $\Delta \omega _1$. Generally, $%
\Delta \omega _1<\overline{\omega }_1$. The same simplification can be taken
for the Lamb phase shift, which is expressed as the mean value of $\lambda
_{2\overrightarrow{k}}$ under the distribution $h_2\left( \omega _{%
\overrightarrow{k}}\right) $. $h_2\left( \omega _{\overrightarrow{k}}\right) 
$ is characterized by $\overline{\omega }_2$ and $\Delta \omega _2$. In the
following we use the four parameters $\overline{\omega }_1,\Delta \omega _1,%
\overline{\omega }_2,\Delta \omega _2$ to discuss the decoherence behavior
of the register. First we show that the cooperative coupling with the
environment can yield independent decoherence of the qubits in certain
circumstances. There are two cases.\\

{\bf Case 1 } $\frac{\overline{\omega }_1\delta }v\geq \pi ,$ $\frac{%
\overline{\omega }_2\delta }v\geq \pi .$

In the above condition, $v$ indicates the velocity of the noise field and $%
\delta $ is the variance of sites of the qubits. Under this condition, for
the effective mode $\overrightarrow{k}$ (a mode $\overrightarrow{k}$ is
called effective if in Eqs. (12) and (13) it has sufficient contributions to
the summation.), $\gamma _{\overrightarrow{l}}=i_{\overrightarrow{l}}e^{i%
\overrightarrow{k}\cdot \overrightarrow{r}_{\overrightarrow{l}}}$ becomes a
random variable which satisfies 
\begin{equation}
\label{20}\left\langle \gamma _{\overrightarrow{l}}\right\rangle =0,\text{ }%
\left\langle \left| \gamma _{\overrightarrow{l}}\right| ^2\right\rangle =1.
\end{equation}
Obviously, the variables $\gamma _{\overrightarrow{l}}$ are independent of
each other. So the mean $\lambda _{2\overrightarrow{k}}$ becomes%
$$
\left\langle \lambda _{2\overrightarrow{k}}\right\rangle =L-L=0 
$$
where $L$ is the total number of the qubits. Similarly, if $i_{%
\overrightarrow{l}}\neq j_{\overrightarrow{l}}$, $\gamma _{\overrightarrow{l}%
}=\frac{i_{\overrightarrow{l}}-j_{\overrightarrow{l}}}2e^{i\overrightarrow{k}%
\cdot \overrightarrow{r}_{\overrightarrow{l}}}$ is also a random variable
satisfying Eq. (20). Suppose $L_0$ is the number of the pairs $\left( i_{%
\overrightarrow{l}},j_{\overrightarrow{l}}\right) $ with $i_{\overrightarrow{%
l}}\neq j_{\overrightarrow{l}}$, the mean $\lambda _{1\overrightarrow{k}}$
is thus simplified to 
\begin{equation}
\label{22}\left\langle \lambda _{1\overrightarrow{k}}\right\rangle =4L_0=%
\stackunder{\overrightarrow{l}}{\sum }\left( i_{\overrightarrow{l}}-j_{%
\overrightarrow{l}}\right) ^2.
\end{equation}
With (21) and (22), the phase damping and the Lamb phase shift become,
respectively, 
\begin{equation}
\label{23}\eta _{\left\{ i_{\overrightarrow{l}},j_{\overrightarrow{l}%
}\right\} }\left( t\right) =x\stackunder{\overrightarrow{l}}{\sum }\left( i_{%
\overrightarrow{l}}-j_{\overrightarrow{l}}\right) ^2,
\end{equation}
\begin{equation}
\label{24}\phi _{\left\{ i_{\overrightarrow{l}},j_{\overrightarrow{l}%
}\right\} }\left( t\right) =0.
\end{equation}
This is just the result of the independent decoherence, which is obtained in
[25] and [26] under the assumption that the qubits interact with different
environments. Here we see, provided the disorder in the register is
sufficiently large, the qubits will be decohered independently, even if they
couple with the same environment. In the independent decoherence, the Lamb
phase shift reduces to zero.\\

{\bf Case 2 } $\frac{\Delta \omega _1d}v>>1,$ $\frac{\Delta \omega _2d}v>>1.$

In the above condition, $d$ indicates the lattice constant. If $%
\overrightarrow{l_1}\neq \overrightarrow{l_2}$, let $\overrightarrow{k}\cdot
\left( \overrightarrow{r}_{\overrightarrow{l_1}}-\overrightarrow{r}_{%
\overrightarrow{l_2}}\right) =\frac{sd}v\omega _{\overrightarrow{k}}$, where 
$s$ and$1$ have the same order of magnitude. Under the distribution $%
h_1\left( \omega _{\overrightarrow{k}}\right) $ or $h_2\left( \omega _{%
\overrightarrow{k}}\right) $, the following mean value 
\begin{equation}
\label{25}\left\langle e^{i\overrightarrow{k}\cdot \left( \overrightarrow{r}%
_{\overrightarrow{l_1}}-\overrightarrow{r}_{\overrightarrow{l_2}}\right)
}\right\rangle =\left\langle e^{i\frac{sd}v\omega _{\overrightarrow{k}%
}}\right\rangle 
\end{equation}
is a Fourier transformation of the weight function. Suppose $\Delta \omega _i
$ is the variance of the distribution, which can be approximated by a
Gaussian function, hence we have 
\begin{equation}
\label{26}\left\langle e^{i\frac{sd}v\omega _{\overrightarrow{k}%
}}\right\rangle \sim \exp \left[ -\left( \frac{\Delta \omega _isd}v\right)
^2\right] \sim 0.
\end{equation}
So after summation over the mode $\overrightarrow{k}$, only the
non-variation terms , such as $\left( i_{\overrightarrow{l}}-j_{%
\overrightarrow{l}}\right) ^2$, in $\lambda _{1\overrightarrow{k}}$ and $%
\lambda _{2\overrightarrow{k}}$ have contributions to the result. We then
obtain Eqs. (23) and (24) again. Therefore, in this case the qubits are also
decohered independently.

To reduce the independent decoherence, many kinds of quantum error
correction schemes have been proposed. However, the two conditions for the
independent decoherence are not always satisfied in practice. In the next
section we discuss other circumstances.

\section{Collective decoherence}

The independent decoherence is an ideal case. In this section, we discuss
another ideal case, the collective decoherence. This requires that the
disorder in the register should be small, i.e., the variance $\delta $
should satisfy $\frac{\overline{\omega }_1\delta }v<<\pi ,$ and $\frac{%
\overline{\omega }_2\delta }v<<\pi $. (We have assumed $\Delta \omega _i\leq 
\overline{\omega }_i$ $\left( i=1,2\right) $.) Under this condition, for the
effective $\overrightarrow{k},$ we approximately have $\overrightarrow{k}%
\cdot \overrightarrow{r}_{\overrightarrow{l}}\approx \overrightarrow{k}\cdot 
\overrightarrow{R}_{\overrightarrow{l}}$ where $\overrightarrow{R}_{%
\overrightarrow{l}}$ is a rigorous periodical function of $\overrightarrow{l}
$. There are two circumstances which can result in the collective
decoherence.\\

{\bf Case 1 } $\frac{\overline{\omega }_1d}v<<\pi ,$ $\frac{\overline{\omega 
}_2d}v<<\pi .$

In this case, two adjacent qubits lie in the coherent length of the
environment. We call two adjacent qubits a qubit-pair. Suppose there are $L$
qubit-pairs (so $2L$ qubits) in the register. The two qubits in the $%
\overrightarrow{l}$ qubit-pair are indicated by $\overrightarrow{l}$ and $%
\overrightarrow{l^{^{\prime }}}$, respectively. Then, for the effective $%
\overrightarrow{k}$, the factor $\lambda _{1\overrightarrow{k}}$
approximately becomes 
\begin{equation}
\label{27}\lambda _{1\overrightarrow{k}}\approx \left| \stackunder{%
\overrightarrow{l}}{\sum }\left( i_{\overrightarrow{l}}+i_{\overrightarrow{%
l^{^{\prime }}}}-j_{\overrightarrow{l}}-j_{\overrightarrow{l^{^{\prime }}}%
}\right) e^{i\overrightarrow{k}\cdot \overrightarrow{R}_{\overrightarrow{l}%
}}\right| ^2.
\end{equation}
$\lambda _{2\overrightarrow{k}\text{ }}$has a similar expression. Eq. (27)
reveals the collective decoherence of the two qubits in a qubit-pair. In the
collective decoherence, the decoherence rate is sensitive to the type of the
input states. The states which undergo no or reduced decoherence are called
''subdecoherent'' states.

The existing quantum error schemes are not suitable for reducing the
collective decoherence. Fortunately, for the collective decoherence, there
is a simpler decoherence-reducing strategy. The input states of $L$ qubits
can be encoded into the ''subdecoherent'' states of $L$ qubit-pairs by the
following encoding 
\begin{equation}
\label{28}
\begin{array}{c}
\left| -1\right\rangle \rightarrow \left| -1,1\right\rangle , \\ 
\left| 1\right\rangle \rightarrow \left| 1,-1\right\rangle .
\end{array}
\end{equation}
Because of Eq. (27) and a similar equation of $\lambda _{2\overrightarrow{k}%
\text{ }}$, the encoded states obviously undergo no phase damping and Lamb
phase shift. So the coherence is preserved. The encoding (28) has been
mentioned in [25] and extensively discussed in [27]. It can be simply
fulfilled by the quantum controlled-NOT gates.\\

{\bf Case 2 } $\frac{\Delta \omega _1Ld}v<<\pi ,$ $\frac{\Delta \omega _2Ld}%
v<<\pi .$

In this case, the lattice constant $d$ and the effective wave length $\frac{%
2\pi v}{\overline{\omega }_i}$ of the noise field have the same order of
magnitude. So the adjacent qubits do not lie in the coherent length of the
environment. But the distribution functions $h_1\left( \omega _{%
\overrightarrow{k}}\right) $ and $h_2\left( \omega _{\overrightarrow{k}%
}\right) $ have a peak and the width of the peak is small so that $Ld<<\frac{%
\pi v}{\Delta \omega _i}$ $\left( i=1,2\right) $. (For simplicity, here we
consider the one-dimensional register. The discussion of the
three-dimensional circumstances is very similar.) From Eq. (18), the phase
damping factor is thus simplified to 
\begin{equation}
\label{29}\eta _{\left\{ i_{\overrightarrow{l}},j_{\overrightarrow{l}%
}\right\} }\left( t\right) =x\lambda _{1\overline{k}_1}=x\left| \stackunder{l%
}{\sum }\left( i_l-j_l\right) e^{i\overline{k}_1R_l}\right| ^2,
\end{equation}
where $\overline{k}_1=\frac{\overline{\omega }_1}v$ and $x$ is given by Eq.
(19). Similarly, the Lamb phase shift $\phi _{\left\{ i_{\overrightarrow{l}%
},j_{\overrightarrow{l}}\right\} }\left( t\right) \propto \lambda _{2%
\overline{k}_2}$. We assume $\overline{k}_2\approx \overline{k}_1=\overline{k%
}.$

Eq. (29) suggests that the qubits are decohered collectively. This results
from the periodicity of the sites $R_l$. Similar to the case 1, the
collective decoherence described by Eq. (29) can also be reduced by pairing
the qubits. But this time the qubit-pairs do not consist of two adjacent
qubits. Since $\frac{\overline{k}d}\pi \sim 1$, there exist round numbers $m$
and $n$ to satisfy $\left| m\frac{\overline{k}d}\pi -n\right| <<1$, where $m$
is chosen as small as possible. Hence we have $e^{i\overline{k}\left(
R_{l+m}-R_l\right) }\approx \left( -1\right) ^n$. So the $l$ qubit and the $%
l+m$ qubit can be put into a pair. The state of the qubits can be
transformed into the ''subdecoherent'' state of the qubit-pairs by the
following encoding 
\begin{equation}
\label{30}
\begin{array}{c}
\left| -1\right\rangle \rightarrow \left| -1,\left( -1\right)
^n\right\rangle , \\ 
\left| 1\right\rangle \rightarrow \left| 1,\left( -1\right)
^{n+1}\right\rangle .
\end{array}
\end{equation}
Obviously the decoherence of the encoded state is reduced.

In the above we require $\frac{\Delta \omega _iLd}v<<\pi $. This condition
is too strong and in fact it is not necessary. It is clear that the
decoherence-reducing strategy described in the above paprgraph still works
if $m\frac{\Delta kd}\pi <<1$, where $m$ is a small round number. So for
reducing this kind of decoherence, we only need $\frac{\Delta \omega _{ii}md}%
v<<\pi $ $\left( i=1,2\right) $.\\

{\bf Acknoledgment}

This project was supported by the National Natural Science Foundation of
China.

\newpage\

\end{document}